\documentclass[%
 pre,
 onecolumn,
superscriptaddress,
nofootinbib,
 amsmath,amssymb,
 aps,
]{revtex4-2}

\usepackage{graphicx}
\usepackage{dcolumn}
\usepackage{bm}
\usepackage[colorlinks=true,citecolor=blue,linkcolor=blue,urlcolor=blue]{hyperref}
\begin{document}

\title{Self-phoretic colloids in chiral active fluids}
\author{Michalis Chatzittofi}
\email{mc2623@cam.ac.uk}
\altaffiliation{These authors contributed equally to this work.}
\affiliation{Max Planck Institute for Dynamics and Self-Organization (MPI-DS), Am Fassberg 17, 37077 G\"{o}ttingen, Germany}
\affiliation{DAMTP, Centre for Mathematical Sciences, University of Cambridge, Cambridge CB3 0WA, United Kingdom}

\author{Yuto Hosaka}
\email{hosaka.yuto.7r@kyoto-u.ac.jp}
\altaffiliation{These authors contributed equally to this work.}
\affiliation{Max Planck Institute for Dynamics and Self-Organization (MPI-DS), Am Fassberg 17, 37077 G\"{o}ttingen, Germany}
\affiliation{Department of Mathematics, Kyoto University, 606-8502 Kyoto, Japan}

\author{Andrej Vilfan}
\affiliation{Max Planck Institute for Dynamics and Self-Organization (MPI-DS), Am Fassberg 17, 37077 G\"{o}ttingen, Germany}
\affiliation{Jo\v{z}ef Stefan Institute, 1000 Ljubljana, Slovenia}

\author{Ramin Golestanian} 
\email{ramin.golestanian@ds.mpg.de}
\altaffiliation{(Corresponding author)}
\affiliation{Max Planck Institute for Dynamics and Self-Organization (MPI-DS), Am Fassberg 17, 37077 G\"{o}ttingen, Germany}
\affiliation{Rudolf Peierls Centre for Theoretical Physics, University of Oxford, Oxford OX1 3PU, UK}
\affiliation{Institute for the Dynamics of Complex Systems, University of G\"{o}ttingen, 37077 G\"{o}ttingen, Germany}

\date{\today}

\begin{abstract}
Autonomous and driven transport in chiral active fluids have been shown to exhibit features that cannot be accommodated within the classical formulation of fluid mechanics, due to the role of odd viscosity. We generalize the theory of phoretic active matter to fluid environments with odd viscosity and derive expressions for translational and rotational self-propulsion velocities in the case of a spherical swimmer with arbitrary activity and mobility surface profiles. We discuss specific examples of chemically active colloids with axisymmetric and non-axisymmetric coatings and the resulting interplay between symmetry and chirality. Our results can be applied to study the emergent collective dynamics of phoretic particles in fluid media with broken time-reversal and parity symmetries.
\end{abstract}

\maketitle

Theoretical and experimental studies of autonomous microscopic swimmers have led to a wealth of discoveries in the last two decades \cite{Chen2025}, which promise a wide range of technological applications \cite{Ju2025}. Moreover, this development has enabled the field of active matter to introduce well-controlled prototypes that can be used for systematic theoretical and experimental studies of self-propelled particles and their emergent collective properties~\cite{Gompper2020Feb}. While living motile systems such as swimming bacteria and crawling cells have been widely used in studies of active matter, synthetic systems that can be made in a uniform and fully predictable manner have played a key role in understanding non-equilibrium phenomena, which can ultimately shed light on the behavior of biological systems. Moreover, they naturally lend themselves to systematic coarse-graining because of the mechanistic understanding of the behavior of the individual self-propelled particles. Among others \cite{Chen2025}, phoretic effects have been found to provide a robust foundation for making self-propelled particles, due to the force-free nature of the interfacial transport mechanisms ~\cite{anderson1989colloid, moran2017phoretic,Golestanian2018phoretic}. Examples include colloids in gradients of chemical concentration (diffusiophoresis)~\cite{howse2007self}, electric field (electrophoresis)~\cite{Ebbens2014Jun}, or temperature field (thermophoresis)~\cite{Sano2010,golestanian2012collective}, and a variety of experimental and theoretical studies have been performed to address different aspects of phoretic propulsion \cite{Paxton2004,FournierBidoz2005,Kapral2007,golestanian2007designing, julicher2009generic, Palacci2013, Bechinger2013b, saha2014clusters,Brown2014,ebbens2016active, zhang2017active,lisicki2018autophoretic}.

To generate self-phoretic motion, the system needs to encompass a certain level of symmetry breaking, as required by the Curie principle \cite{Golestanian2018phoretic}, typically by introducing asymmetry in the surface properties~\cite{golestanian2007designing}. For example, a Janus particle with a half-coated spherical cap is commonly used in experiments, with self-propulsion arising due to its fore-aft asymmetry \cite{howse2007self}. While the combination of shape asymmetry and phoretic activity can lead to the emergence of spontaneous spin \cite{Ebbens2010,Bechinger2013,Soto2014}, it is well known that Janus spheres cannot spontaneously rotate in the bulk due to their axisymmetric surface properties. However, the presence of external symmetry breaking mechanisms such as gradients in the external field~\cite{Tatulea-Codrean2018Dec}, rigid walls~\cite{Uspal2015,das2015boundaries, uspal16,mozaffari16,Ibrahim2016Oct, yariv2017boundary, reigh2018diffusiophoretically, daddi2022diffusiophoretic}, or shear flows~\cite{premlata2021coupled}, can lead to the emergence of rotation.
Geometric asymmetry in the particle shape~\cite{ibrahim2018shape, poehnl2020axisymmetric, popescu2011pulling, michelin2015autophoretic} or even spontaneous symmetry breaking~\cite{michelin2013spontaneous, hu2022spontaneous,brandao2023spontaneous,Wittmann2025} have also been shown to induce self-phoretic locomotion.
In this article, we address the topic of phoretic activity in the presence of chiral symmetry breaking in the bulk, and whether odd viscosity can change the symmetry requirements for the emergence of translational and rotational self-propulsion.

We consider the case in which a phoretic colloidal particle moves in a 3D chiral active fluid with broken time-reversal and parity symmetries. Employing the Stokes equation with added odd viscosity as the continuum description of chiral active fluids and the generalized Lorentz reciprocal theorem, we derive expressions for the translational and rotational velocities a phoretic sphere with arbitrary surface coating of activity and mobility. We show that translational velocity remains the same as in the case without odd viscosity, whereas the rotational velocity is affected by odd viscosity. In particular, a Janus particle undergoes the orientational dynamics in the bulk, which would be forbidden by symmetry in classical fluids.
Our findings uncover a transport mechanism in aqueous environments characterized by chirality or parity violation. 

\begin{figure}
	\centering
	\includegraphics[width=.5\linewidth]{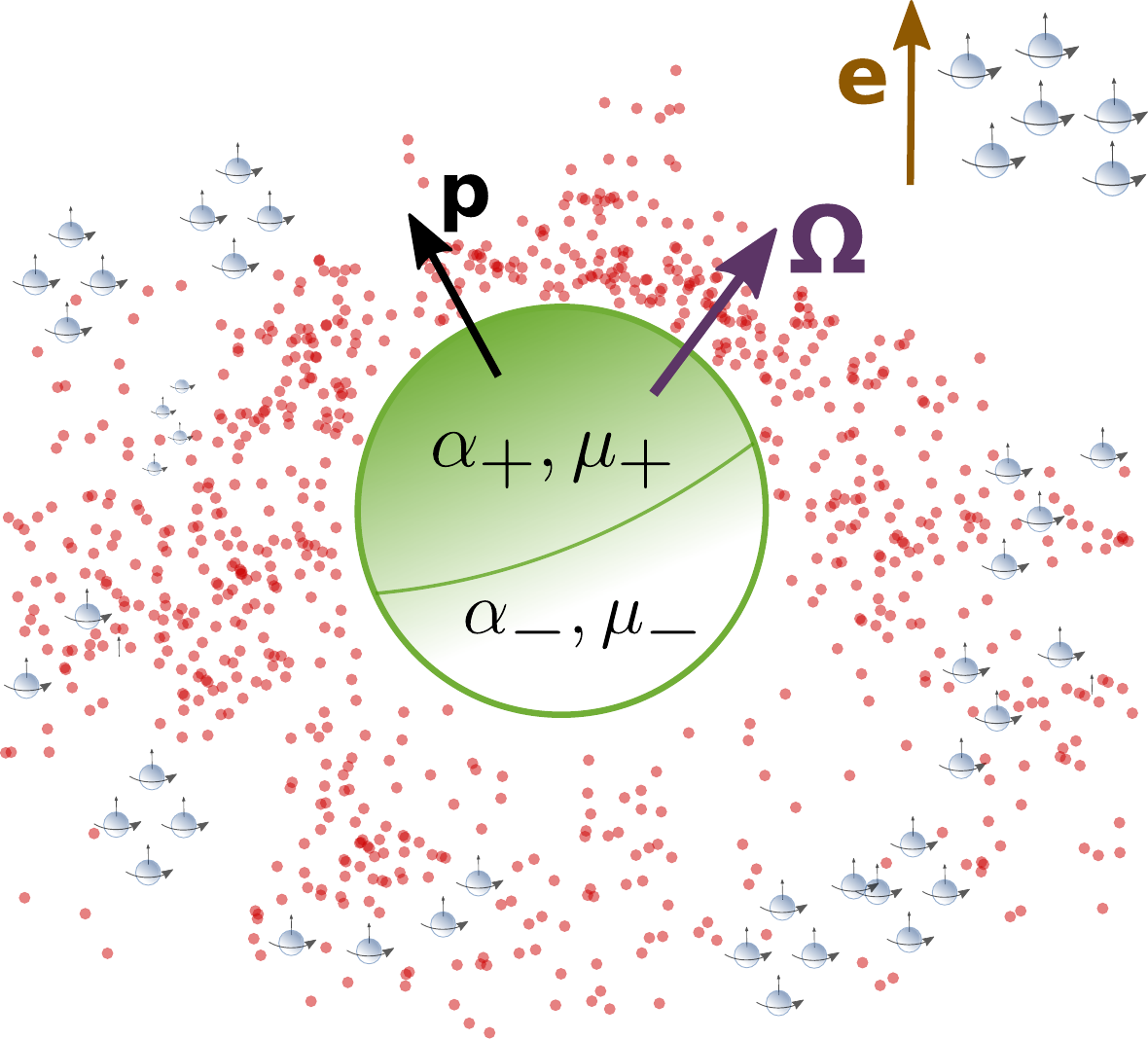}
	\caption{
        Schematic of a self-phoretic Janus particle in a 3D chiral active fluid with even (shear) viscosity $\eta^{\rm e}$ and odd viscosity $\eta^{\rm o}$.
        The Janus swimmer of radius $a$ with a broken fore-aft symmetry in the mobility $\mu$ and the activity $\alpha$ moves with translational velocity $\mathbf{V}=V\mathbf{p}$ and rotational velocity $\boldsymbol{\Omega}$, the latter arising from the odd viscosity.
        The unit vector $\mathbf{e}$ denotes the axis of chirality of the surrounding fluid. Microscopically, its direction can correspond to the axis around which microscopic fluid components spin.
        }
	\label{fig:intro}
\end{figure}

\section*{3D fluids with odd viscosity}

In recent years, various theoretical developments have been made in the direction of transport phenomena in fluids with odd viscosity~\cite{avron1998odd, hosaka2022nonequilibrium, fruchart2023odd}.
More recently, the behavior of passive bodies in motion has been investigated in the literature and found to deviate from classical results in fluid mechanics. Examples include the flow fields past a sphere or a disk~\cite{khain2022stokes, Lier_2024, hosaka2021hydrodynamic} as well as the resulting anti-symmetric components in the Green function~\cite{hosaka2021nonreciprocal, khain2022stokes, yuan2023stokesian, everts2024dissipative} and the resistance tensor of a spherical body~\cite{hosaka2023lorentz, everts2024dissipative, hosaka2024chirotactic}.
These asymmetries naturally lead to a \textit{non-reciprocal} coupling between forces and velocities, as for instance recently demonstrated in the case of an active dimer~\cite{w6pg-4471}. Notably, they result from a violation of Onsager reciprocity as a direct consequence of how odd viscosity affects the motion of particles.
Furthermore, to study autonomous transport of active particles, the Lorentz reciprocal theorem has been generalized to account for odd viscosity and enable the calculation of the velocity of microswimmers in chiral active fluids~\cite{hosaka2023lorentz}.
Other approaches using the linear response Green's function~\cite{hosaka2023pair, hosaka2023hydrodynamics} or the geometric theory~\cite{lapa2014} can also be used in studying swimming at low Reynolds number.

We consider an incompressible Stokes flow with odd viscosity in 3D as a model to represent chiral active fluids~\cite{markovich2021odd, khain2022stokes}.
The 3D \textit{odd} Stokes equations are governed by the force balance equation $\partial_j\sigma_{ij}=0$ and the continuity condition ${\bm \nabla}\cdot\mathbf{v}=0$. The stress tensor is given as $\sigma_{ij}=-p\delta_{ij}+\eta_{ijk\ell}\partial_\ell v_k$ in terms of the velocity field $\mathbf{v}$, the pressure $p$, and the 4th-rank viscosity tensor $\boldsymbol{\eta}$.
Despite having 81 possible components in general, $\boldsymbol{\eta}$ can be expressed simply with two viscosity coefficients via the symmetry argument presented below. In the simplest case of an isotropic continuum, $\boldsymbol{\eta}$ is characterized by a single coefficient: the even (shear) viscosity, $\eta^{\rm e}$. 
However, since there are no odd viscosities compatible with spatial isotropy in 3D, chiral fluids must have at least one preferred axis~\cite{khain2022stokes, fruchart2023odd}.
One of the simplest continuum media that features chirality is the one having the cylindrical symmetry about that direction with reflection symmetry across the plane perpendicular to it~\cite{khain2022stokes}.
Such a system can be realized by fluids composed of microscopic particles spinning along one axis (Fig.~\ref{fig:intro}).
Imposing these constraints and assuming that the resulting odd viscosities are proportional to each other (as assumed in previous studies~\cite{khain2022stokes, yuan2023stokesian, everts2024dissipative, Khain2024Aug, Lier_2024}) reveals the following symmetry properties of the whole viscosity tensor; the components represented with $\eta^{\rm e}$ are even under the transformation ${ij}\leftrightarrow{k\ell}$, while those represented with the odd viscosity, $\eta^{\rm o}$, are odd.
Denoting the axis of chirality by $\mathbf{e}$, the stress tensor of such a 3D fluid can be written as 
\begin{align}
    \sigma_{ij}
    =
    -p\delta_{ij}+2\eta^{\rm e}E_{ij}
    +
    \eta^{\rm o}
    e_\ell
    (
    \epsilon_{\ell ik}E_{kj}
    +
    \epsilon_{\ell jk}E_{ki}
    )
    ,
\label{eq:stress}
\end{align}
where $E_{ij}=\frac{1}{2}(\partial_iv_j+\partial_jv_i)$ is the strain-rate tensor, and $\epsilon_{ijk}$ is the 3D Levi-Civita tensor.
Since the antisymmetric components of $\boldsymbol{\eta}$ are not associated with fluid dissipation, $\eta^{\rm o}$ can be either positive or negative, while the even counterpart is strictly positive $(\eta^{\rm e}>0)$, as required by the positivity of the entropy production.

\section*{Swimming dynamics}

The Lorentz reciprocal theorem is a powerful tool for studying low-Reynolds-number fluid dynamics, which provides an integral identity linking a main problem to be solved with a commonly known auxiliary problem~\cite{masoud}.
By using the theorem with the classical solution for a moving sphere as the auxiliary problem, swimming speeds of surface-driven microswimmers~\cite{stone1996propulsion} and self-phoretic particles~\cite{golestanian2007designing} have been derived without directly solving the main swimming problem.
Some of the authors have recently shown that the theorem can be generalized to fluids with odd viscosity~\cite{hosaka2023lorentz}, by reversing the sign of the odd viscosity between the main and auxiliary problem such that $\eta^{\rm o} = -\hat \eta^{\rm o}$, while the even viscosities are kept equal, i.e., $\eta^{\rm e} =\hat \eta^{\rm e}$. 
Here, the hat symbol corresponds to the auxiliary solution.

We utilize the generalized Lorentz reciprocal theorem to determine the translational and rotational velocities of a spherical active particle of radius $a$ in fluids with odd viscosity. 
Suppose the surface-driven swimmer with a prescribed effective slip velocity $\mathbf{v}^{\rm s}$ in a fluid characterized by the odd-to-even viscosity ratio $\lambda=\eta^{\rm o}/\eta^{\rm e}$ as the main and the dragged sphere as the auxiliary problem.
The reciprocal theorem can be written in the form~\cite{stone1996propulsion, hosaka2023lorentz}
\begin{align}
    \mathbf{V}\cdot\hat{\mathbf{F}}
    +
    \boldsymbol{\Omega}\cdot\hat{\mathbf{T}}
    &=
    -\int_\mathcal{S}dS\, \mathbf{v}^{\rm s}\cdot\hat{\mathbf{f}},
    \label{eq:samuelV}
\end{align}
where $\mathbf{V}$ and $\boldsymbol{\Omega}$ are the translational and angular velocity of a force- and torque-free swimmer $\mathbf{F}=\mathbf{T}=\mathbf{0}$, $\hat{\mathbf{F}}$ and $\hat{\mathbf{T}}$ are the force and torque on the passive sphere, respectively, and $\hat{\mathbf{f}}=\hat{\boldsymbol{\sigma}}\cdot\mathbf{n}$ is its surface traction with $\mathbf{n}$ being a surface normal pointing into the fluid.

To obtain the velocity of a spherical microswimmer, one needs to derive the surface traction $\hat{\mathbf{f}}$ and net force $\hat{\mathbf{F}}$ on a translating sphere of the same size.
Although $\hat{\mathbf{f}}$ and $\hat{\mathbf{F}}$ depend on the odd viscosity in a very complicated way~\cite{everts2024dissipative}, one can find the relation between these quantities in a surprisingly simple form $\hat{\mathbf{F}}=4\pi a^2\hat{\mathbf{f}}$ with the radius $a$.
This is because the solution to this auxiliary problem assumes constant surface traction, and this assumption is verified when the boundary condition is satisfied.
By applying the constant traction to the Lorentz reciprocal theorem~\eqref{eq:samuelV}, the swimming velocity follows  
\begin{align}
    \mathbf{V} = - \frac{1}{4\pi a^2}\int_\mathcal{S}dS\, \mathbf{v}^{\rm s}
    ,
    \label{eq:V}
\end{align}
which is independent of the odd viscosity, as consistent with the classical result~\cite{stone1996propulsion}.
The translational velocity is therefore not affected by any value of the odd viscosity, extending the results previously derived to 
linear order in $\lambda$~\cite{hosaka2023lorentz} and for swimmers with the leading squirming mode (source dipole)~\cite{hosaka2024chirotactic}.

The angular velocity of a spherical microswimmer in a chiral fluid in 3D has been analytically studied by the present authors in a previous communication~\cite{hosaka2024chirotactic}.
For a rigidly rotating sphere, the exact expressions of the traction and the torque are given by~\cite{hosaka2024chirotactic}
\begin{align}
    \hat{\mathbf{f}}
    &=
    -3
    \hat{{\eta}}^{\rm 
    e}
    \left(
    \hat{\boldsymbol{\Omega}}\times
    \mathbf{n}
    +
    \frac{\hat{\lambda}}{2}
    (\hat{\boldsymbol{\Omega}}\times\mathbf{n})\times\mathbf{e}
    +
    \frac{\hat{\lambda}}{3}
    (\hat{\boldsymbol{\Omega}}\cdot\mathbf{e})
    \mathbf{n}
    \right),\\
    \hat{\mathbf{T}} 
    &=
    -8\pi\hat{\eta}^{\rm e}a^3
    \left(
    \hat{\boldsymbol{\Omega}}
    +
    \frac{\hat{\lambda}}{4}
    \hat{\boldsymbol{\Omega}}\times\mathbf{e}
    \right).
    \label{eq:torque}
\end{align}
Applying the generalized Lorentz reciprocal theorem~\eqref{eq:samuelV} into these expressions gives the angular velocity $\boldsymbol{\Omega}$ for an arbitrary odd-to-even viscosity ratio, $\lambda$~\cite{hosaka2024chirotactic}.
The angular velocity can be decomposed into two contributions as follows $\boldsymbol{\Omega}=\boldsymbol{\Omega}^{\rm e}+\boldsymbol{\Omega}^{\rm o}$; the (even) angular velocity $\boldsymbol{\Omega}^\mathrm{e}$ is purely due to the intrinsic rotation and $\boldsymbol{\Omega}^\mathrm{o}$ is induced by the odd viscosity~\cite{hosaka2024chirotactic}. We find
\begin{align}
    \boldsymbol{\Omega}^{\rm e}
    &=
    -\frac{3}{8\pi a^3}
    \int_\mathcal{S}dS\,\mathbf{n}\times\mathbf{v}^{\rm s},
    \label{eq:Omegae}
    \\
    \boldsymbol{\Omega}^{\rm o}
    &=
    \frac{3}{20\pi a^3}\,
    \mathbf{e}\cdot
    \int_\mathcal{S}dS\,
    \left(-\frac{5}{2}(\mathbf{v}^{\rm s}\mathbf{n}+
    \mathbf{n}\mathbf{v}^{\rm s})+(\mathbf{n}\cdot\mathbf{v}^{\rm s})\mathbf{I}\right)
    \cdot
    \left(
    \frac{\lambda}{4}\,
    \mathbf{I}
    +
    \frac{(\lambda/4)^2}{1+(\lambda/4)^2}
    \boldsymbol{\epsilon}\cdot\mathbf{e}
    +
    \frac{(\lambda/4)^3}{1+(\lambda/4)^2}
    (\mathbf{ee}-\mathbf{I})\right)
    ,
    \label{eq:Omega}
\end{align}
where we have assumed volume conservation within the swimmer $(\int_\mathcal{S}dS\,\mathbf{v}^{\rm s}\cdot\mathbf{n}=0)$.
While $\boldsymbol{\Omega}^{\rm e}$ coincides with the classical expression~\cite{stone1996propulsion}, $\boldsymbol{\Omega}^{\rm o}$ exhibits the nonlinear dependence on $\lambda$ and has a non-trivial combination of the slip velocity in the surface integral, which includes the stresslet or force-dipole moment~\cite{hosaka2024chirotactic, lauga2016stresslets, nasouri2018higher}.
Equations~\eqref{eq:V}, \eqref{eq:Omegae}, and \eqref{eq:Omega} represent our main foundation and can be applied to any slip velocity profile $\mathbf{v}^{\rm s}$ in 3D fluids with odd viscosity.
We note the contrast with the 2D case, where there is no modification to translational and rotational velocities in 2D fluids with odd viscosity~\cite{hosaka2024chirotactic, lapa2014}.

\section*{Stresslet induced by self-phoretic particles with arbitrary coverage}

Having obtained the swimming velocity induced by the effective surface velocity, we proceed to derive a general expression for the velocity of a self-phoretic spherical particle having arbitrarily catalytic patches.
In doing so, we expand the local activity and the mobility of the particle in spherical coordinates $(r,\theta,\phi)$ in the swimmer-fixed frame where the swimming direction $\mathbf{p}$ is set to be the positive polar axis. 
Since our focus in this study is on the $\lambda$-dependent swimming velocity, we determine only the odd rotational velocity $\boldsymbol{\Omega}^{\rm o}$ in terms of expansion coefficients, while $\mathbf{V}$ and $\boldsymbol{\Omega}^{\rm e}$ for general phoretic particles were derived before~\cite{golestanian2007designing,lisicki2018autophoretic}.
From the observation of Eq.~\eqref{eq:Omega}, the current problem can be reduced to the calculation of the stresslet of a phoretic sphere with tangential slip velocity alone
\begin{align}
    \mathbf{S}
    =
    -
    \frac{5}{16\pi}
    \int_\mathcal{S}dS\,
    (\mathbf{v}^{\rm s}\mathbf{n}+
    \mathbf{n}\mathbf{v}^{\rm s})
    .
    \label{eq:stresslet}
\end{align}
Stresslets induced by axisymmetric phoretic swimmers have been studied previously~\cite{lauga2016stresslets, poehnl2020axisymmetric}.

The concentration of the solute particles is described by the scalar field $C$ whose dynamics is governed by the diffusion equation.
Interactions between the particle surface and the solute lead to an effective surface velocity tangential to the surface, which is proportional to the local concentration gradient along the surface, namely
\begin{align}
    \mathbf{v}^{\rm s} = \mu(\theta,\phi) (\mathbf{I}-\mathbf{nn})\cdot{\bm \nabla} C
    ,
    \label{eq:vs}
\end{align}
where $\mu(\theta,\phi)$ is the surface mobility and we have assumed that the slip velocity does not include any contributions due to odd viscosity.
Microscopically, such a regime can be obtained by the separation of length scales between the size of the swimmers and that of the mesoscopic constituents of the chiral active fluid that give rise to odd viscosity (e.g., spinning particles)~\cite{markovich2021odd}.
Moreover, the self-phoretic particle has a fixed non-equilibrium flux due to the activity on its surface described by $\alpha(\theta,\phi)$, which acts as a sink or source for solutes.
When the solute diffusion dominates over advection (in the low P\'{e}clet number regime), the concentration field can be obtained as the solution to the Laplace equation~\cite{golestanian2007designing}
\begin{align}
    {\bm \nabla}^2 C =0,
\end{align}
subjected to the following boundary conditions
\begin{align}
    -D \mathbf{n} \cdot {\bm \nabla} C|_{r=a}= \alpha(\theta,\phi),\quad
    \lim_{r\to \infty}C(r,\theta,\phi)=0,
\end{align}
where $D$ is the diffusion coefficient of the solute particles.
By expanding the activity and mobility distributions in spherical harmonics as 
\begin{align}
    \alpha(\theta, \phi)=\sum_{\ell m} A_{\ell m} Y_{\ell m}(\theta, \phi), \\
\mu(\theta, \phi)=\sum_{\ell m} M_{\ell m} Y_{\ell m}(\theta, \phi), 
\end{align}
the solution for the concentration profile is obtained as 
\begin{align}
    C(r,\theta,\phi)=\sum_{\ell=0}^{\infty} \sum_{m=-\ell}^{\ell}\frac{A_{\ell m} a^{\ell+2}}{(\ell+1) D} \,\frac{Y_{\ell m}(\theta, \phi)}{r^{\ell+1} },
\end{align}
where the spherical harmonic of degree $\ell$ and azimuthal index $m$ is defined as $Y_{\ell m}(\theta, \phi)=N_{\ell m} P_{\ell}^{m}\mathrm{e}^{i m \phi}$ with the constant prefactor given as $N_{\ell m}=\sqrt{\frac{(2 \ell+1)}{4\pi} \frac{(\ell-m)!}{(\ell+m)!}}$ and $P_{\ell}^{m}\equiv P_{\ell}^{m}(\zeta)$ representing the associated Legendre polynomials $(\zeta \equiv \cos\theta)$. The $Y_{\ell m}$ functions form an orthonormal basis, namely, $\int d\Omega\,Y_{\ell m} Y^*_{\ell'm'} = \delta_{\ell \ell'} \delta_{m m'}$ where $\Omega$ represents the solid angle.

Defining the rescaled activity and mobility coefficients as $\alpha_{\ell m}=N_{\ell m} A_{\ell m}$ and $\mu_{\ell m}=N_{\ell m} M_{\ell m}$, the stresslet of a sphere in Eq.~\eqref{eq:stresslet} can now be evaluated as
\begin{align}
    \mathbf{S} = 
    S\left(\mathbf{pp}-\frac{1}{3}\mathbf{I}\right)
    +
    \frac{5a^2}{16D}
    \sum_{\ell m \ell^\prime}
    \frac{\alpha_{\ell m}}{\ell+1}
    \int_{-1}^1
    d\zeta\left(
    \frac{\partial P_\ell^m}{\partial\zeta}\sqrt{1-\zeta^2}
    \mathbf{B}_{\ell^\prime  m}
    +
    i m P_\ell^m\mathbf{D}_{\ell^\prime  m}
    \right)
    .
    \label{eq:phS}
\end{align}
The above expression applies to the general case of a phoretic sphere with any structure for the activity and mobility. The stresslet tensor $\mathbf{S}$ contains two contributions: the first term is the axisymmetric dipole moment, which is invariant by rotation around the swimming direction $\mathbf{p}$, and the second term is orthogonal to it, as the term vanishes in the quadratic form $\mathbf{p}\cdot\mathbf{B}_{\ell^\prime  m}\cdot\mathbf{p}=\mathbf{p}\cdot\mathbf{D}_{\ell^\prime  m}\cdot\mathbf{p}=0$.
Defining the spectral stresslet coefficients $S_{\ell m}$ through $S=\sum_{\ell m}S_{\ell m}$, we obtain
\begin{align}
    S_{\ell m}
    =&
    \frac{15a^2}{4D}
    \frac{(-1)^{m+1}\alpha_{\ell m}}{(\ell+1)(2\ell+1)}\nonumber\\
    &\times\Biggl[
    \frac{(\ell+1)(\ell+m-1) (\ell+m)}{(2\ell-3)(2\ell-1) }\mu_{\ell-2,-m}
    +
   \frac{\ell^2+\ell-3 m^2}{(2 \ell-1) (2\ell+3)} \mu_{\ell,-m}
    -
    \frac{\ell (\ell-m+1) (\ell-m+2)}{(2\ell+3)(2\ell+5) }
    \mu_{\ell+2,-m}
    \Biggr].
\end{align}
In the second term in Eq.~\eqref{eq:phS}, the tensorial coefficients with the body-fixed frame $(\mathbf{e}_1,\mathbf{e}_2,\mathbf{e}_3)$ with $\mathbf{e}_3\equiv\mathbf{p}$ are given by
\begin{align}
    \mathbf{B}_{\ell^\prime  m}
    =&
    \zeta\sqrt{1-\zeta^2}
    \left[\Delta_2^+(\ell^\prime,m)(\mathbf{e}_1\mathbf{e}_1-\mathbf{e}_2\mathbf{e}_2)
    -
    i\Delta_2^-(\ell^\prime,m)(\mathbf{e}_1\mathbf{e}_2+\mathbf{e}_2\mathbf{e}_1)
    \right]\nonumber\\
    &+
    (2\zeta^2-1)
    \left[
    \Delta_1^+(\ell^\prime,m)(\mathbf{e}_1\mathbf{p}+\mathbf{p}\mathbf{e}_1)
    -i\Delta_1^-(\ell^\prime,m)(\mathbf{e}_2\mathbf{p}+\mathbf{p}\mathbf{e}_2)
    \right],
    \label{eq:Blm}
    \\
    \mathbf{D}_{\ell^\prime  m}
    =&
    -i\Delta_2^-(\ell^\prime,m) (\mathbf{e}_1\mathbf{e}_1-\mathbf{e}_2\mathbf{e}_2)
    -
    \Delta_2^+(\ell^\prime,m)
    (\mathbf{e}_1\mathbf{e}_2+\mathbf{e}_2\mathbf{e}_1)\nonumber\\
    &-
    \frac{\zeta}{\sqrt{1-\zeta^2}}
    \left[
    i
        \Delta_1^-(\ell^\prime,m)
    (\mathbf{e}_1\mathbf{p}+\mathbf{p}\mathbf{e}_1)
    +
        \Delta_1^+(\ell^\prime,m) 
    (\mathbf{e}_2\mathbf{p}+\mathbf{p}\mathbf{e}_2)  
    \right],
    \label{eq:Dlm}
\end{align}
with $\Delta_n^\pm (\ell,m)=\mu_{\ell, -(m+n)}P_{\ell}^{-(m+n)}\pm \mu_{\ell, -(m-n)}P_{\ell}^{-(m-n)}$.
It is more straightforward to perform the remaining integration over $\zeta$ in Eq.~\eqref{eq:phS} numerically rather than seeking a closed-form analytical expression, except for cases where there are a finite number of terms in the spectral expansion.

\section*{Axisymmetric phoretic mobility and activity}

Having obtained the stresslet for arbitrary surface activity and mobility profiles expanded in the spherical harmonics with coefficients $\alpha_{\ell m}$ and $\mu_{\ell m}$, we first study the angular velocity of an axisymmetric phoretic particle. In the case of vanishing azimuthal modes $(m=0)$, symmetry requires $\mathbf{\Omega}^{\rm e}=\mathbf{0}$, while the odd contribution $\mathbf{\Omega}^{\rm o}$ becomes nonzero because of its stresslet moment.
Noting that all the azimuthal coefficients vanish, i.e., $\alpha_{\ell m \neq 0} = 0$ and $\mu_{\ell m \neq 0} =0$, we find $\mathbf{B}_{\ell^\prime0}=\mathbf{D}_{\ell^\prime0}=\mathbf{0}$.
The stresslet can then be obtained as $\mathbf{S}=S(\mathbf{pp}-\frac{1}{3}\mathbf{I})$ with strength
\begin{align}
    S =
    \sum_\ell S_{\ell0}
    =
    -
    \frac{15a^2}{4D}
    \sum_\ell
    \frac{\ell \alpha_{\ell}}{2\ell+1}
    \left[ 
    \frac{\ell-1}{(2\ell-3)(2\ell-1)}
    \mu_{\ell-2}
    +
    \frac{1}{(2\ell-1)(2\ell+3)}\mu_{\ell}
    -
    \frac{\ell+2}{(2\ell+3)(2\ell+5)}\mu_{\ell+2}
    \right],
    \label{eq:Saxi}
\end{align}
where we have defined $\alpha_{\ell}\equiv  \, \sqrt{\frac{2\ell+1}{4\pi}} \, A_{\ell0}$ and $\mu_{\ell }\equiv \, \sqrt{\frac{2\ell+1}{4\pi}} \, M_{\ell0}$, in terms of the coefficients expanded in terms of Legendre polynomials for axially symmetric profiles \cite{golestanian2007designing}.
The sign of the stresslet magnitude distinguishes between the types of microswimmers: $S>0$ corresponds to pullers, which are pulled forward, while $S<0$ corresponds to pushers, which are pushed from behind~\cite{lauga2020fluid}.

In order for swimmers to have a nonvanishing translational velocity, a certain level of symmetry breaking is required in the mobility and activity for the propulsion, as manifested by the Curie principle~\cite{Golestanian2018phoretic}.
In contrast to this requirement, Eq.~\eqref{eq:Saxi} reveals that self-rotation can emerge when both modes have the same parity due to the broken symmetry in the surrounding fluids.
This symmetry argument suggests that the stresslet moment can contribute to motion in odd fluids, unlike in standard fluids where stresslets decouple from self-propulsion due to their fore-aft symmetry and only generate a bulk dipolar flow in the environment.

We now consider specific surface profiles to demonstrate example coverage patterns which result in the rotation of a swimmer with axially symmetric catalytic coatings alone.

\subsection{Saturn particle}

\begin{figure}[t]
	\centering
\includegraphics[width=.5\linewidth]{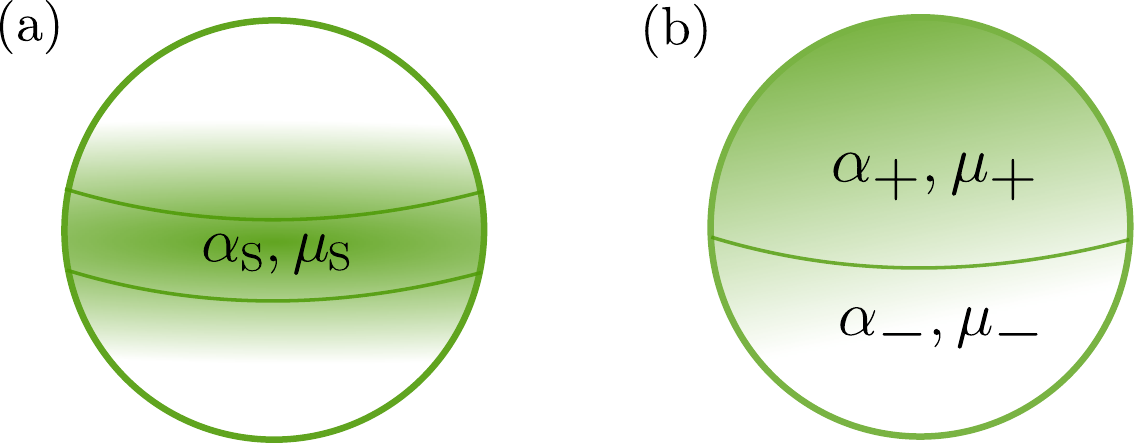}
	\caption{
        Two examples of spherical active phoretic colloids with axisymmetric coating 
        (a) Saturn and (b) Janus \cite{golestanian2007designing}.}
	\label{fig:fig2}
\end{figure}

We first consider the so-called Saturn swimmer [Fig.~\ref{fig:fig2}(a)], whose surface activity is concentrated around the equator $(\theta=\pi/2)$ while the mobility is uniformly distributed on the surface~\cite{golestanian2007designing, katsamba2022chemically}. 
For analytical tractability, we solve the following very similar problem 
\begin{align}
    \alpha(\theta)=\alpha_{\rm s}(1-\cos^2\theta)=\alpha_0 P_0+\alpha_2 P_2(\cos\theta),\quad
    \mu(\theta)=\mu_\mathrm{s},
\end{align}
where $P_\ell(\cos \theta)$ are the Legendre polynomials, and the coefficients are related as $\alpha_0=\frac{2}{3} \alpha_{\rm s}$, $\alpha_2=-\frac{2}{3} \alpha_{\rm s}$, and $\mu_0 = \mu_\mathrm{s}$. 
Since the profile has fore-aft symmetry, both translational and rotational velocities without odd viscosity vanish $(\mathbf{V}=\boldsymbol{\Omega}^{\rm e}=\mathbf{0})$~\cite{golestanian2007designing,lisicki2018autophoretic}. However, the dipolar activity allows for nonvanishing $\boldsymbol{\Omega}^{\rm o}$.
Inserting the axisymmetric stresslet (\ref{eq:Saxi}) into the odd angular velocity (\ref{eq:Omega}) gives
\begin{align}
    \boldsymbol{\Omega}=
    \frac{2\alpha_{\rm s} \mu_\mathrm{s}}{5aD} \,
    \mathbf{e}\cdot\left(\mathbf{pp}-\frac{1}{3}\mathbf{I}\right)\cdot\mathbf{G}(\lambda),
\end{align}
where the scaling function of $\lambda$ is 
\begin{align}
    \mathbf{G}(\lambda)
    =
    \frac{\lambda}{4}\,
    \mathbf{I}
    +
    \frac{(\lambda/4)^2}{1+(\lambda/4)^2}
    \boldsymbol{\epsilon}\cdot\mathbf{e}
    +
    \frac{(\lambda/4)^3}{1+(\lambda/4)^2}
    (\mathbf{ee}-\mathbf{I}).
\end{align}
When the axis of odd viscosity is perpendicular to the swimming direction, i.e., $\mathbf{e}\cdot\mathbf{p}=0$, the swimmer exhibits only spinning motion around $\mathbf{e}$, which is clockwise for $\alpha_{\rm s} \mu_\mathrm{s}\lambda>0$ and anti-clockwise for $\alpha_{\rm s} \mu_\mathrm{s}\lambda<0$.
Note that uniform activity (represented by $\alpha_0$) does not give rise to any rotation, because in that case the tangential gradient of $C$ (and, hence, the slip velocity) vanishes.

\subsection{Janus particle}

A commonly used strategy to create active particles involves different uniform surface profiles with discrete jumps between them, such as the Janus swimmer \cite{howse2007self,golestanian2007designing}, which has the following activity and mobility profiles [Fig.~\ref{fig:fig2}(b)]
\begin{align}
      (\alpha(\theta), \mu(\theta) )=\begin{cases} 
      (\alpha_+,\mu_+), &  0 \leq \theta\leq \frac{\pi}{2}, \\
      (\alpha_-,\mu_-), & \frac{\pi}{2} \leq \theta\leq \pi.
   \end{cases}   
\end{align}
The coating is expanded in the basis of Legendre polynomials as follows
\begin{align}
    \alpha(\theta) &= \sum_\ell \alpha_\ell P_\ell = \frac{1}{2} (\alpha_+ +\alpha_-) P_0 + (\alpha_+ - \alpha_-)\sum_{j=1}^\infty (-1)^{j+1} \frac{4j-1}{j 2^{2j}} {2j-2\choose j-1} P_{2j-1},\label{eq:ALeg}\\
    \mu(\theta) &= \sum_\ell \mu_\ell P_\ell = \frac{1}{2} (\mu_+ +\mu_-) P_0 + (\mu_+ - \mu_-)\sum_{j=1}^\infty (-1)^{j+1} \frac{4j-1}{j 2^{2j}} {2j-2\choose j-1} P_{2j-1}\label{eq:MLeg}.
\end{align}
In this case, Janus swimmer moves with translational velocity 
$\mathbf{V} =-\frac{1}{8D}(\alpha_+ - \alpha_-) (\mu_+ + \mu_-)\mathbf{p}$~\cite{golestanian2007designing}.
The stresslet moment, which is usually irrelevant to self-propulsion, leads to active rotation in chiral systems.
Because of the discrete jump in the coverage at the equator, only the odd degrees of the Legendre polynomials survive in Eq.~\eqref{eq:Saxi}. The net angular velocity of a Janus particle then becomes
\begin{align}
    \boldsymbol{\Omega}
    &=-
    \frac{c_0}{aD}
    (\alpha_+ - \alpha_-) (\mu_+-\mu_-)\,
    \mathbf{e}\cdot \left(\mathbf{pp}-\frac{1}{3}\mathbf{I}\right)\cdot \mathbf{G}(\lambda),
\end{align}
where
\begin{align}
    c_0 &= 
    \frac{9}{2} \sum_{j=1} \frac{2j-1}{j2^{4j-2}}{2j-2\choose j-1} 
    \left[
    \frac{-(2j-2)}{(j-1)(4j-3)}{2j-4\choose j-2}
    + \frac{4j-1}{4j (4j-3)(4j+1)}{2j-2\choose j-1} 
    + \frac{2j+1}{16(j+1)(4j+1)}{2j\choose j}\right]
    \notag
    \\
    &\simeq 0.131. 
\end{align}
We note that for $j=1$ the first term in the square brackets is zero.
Whereas translational propulsion only requires the activity to be different between the two hemispheres, we find that for angular propulsion the symmetry needs to be broken
in both mobility and activity ($\alpha_+\neq \alpha_-$ and $\mu_+\neq \mu_-$). This condition is naturally fulfilled for Janus particles due to the difference in the material properties of the underlying plastic colloids and the metallic catalytic coating \cite{howse2007self}.

It has been shown that an axially symmetric swimming body undergoes the rich orientational dynamics including self-spinning, precession, and reorientation in fluids with odd viscosity~\cite{hosaka2024chirotactic}.
In particular, the reorientation dynamics, in which swimmers show parallel or anti-parallel alignment along the axis of odd viscosity $\mathbf{e}$, is a signature of emerging properties in 3D chiral fluids, termed \textit{bimodal chirotaxis}~\cite{hosaka2024chirotactic}.
Here we discuss the trajectories of a Janus sphere endowed with the bimodal chirotactic response, which depend on the coating properties of catalytic patches and the angle between the swimming direction and the axis of chirality.
For $(\alpha_+ - \alpha_-) (\mu_+-\mu_-)<0$, a Janus swimmer as a puller tends to align with the plane perpendicular to the $\mathbf{e}$-axis and eventually moves on circular paths with radius
$d = \frac{3a}{2c_0|\lambda|}|(\mu_+ + \mu_-)(\mu_+ - \mu_-)^{-1}|.$
For $(\alpha_+ - \alpha_-)(\mu_+ - \mu_-)>0$, the swimmer as a pusher approaches in the direction $\pm\mathbf{e}$, followed by a helical motion ultimately having pitch of the spinning motion $h = \frac{3\pi a}{2c_0|\lambda|}|(\mu_+ + \mu_-)(\mu_+ - \mu_-)^{-1}|$.

\section*{Non-axisymmetric phoretic mobility and activity}

\begin{figure}[h]
	\centering
\includegraphics[width=.8\linewidth]{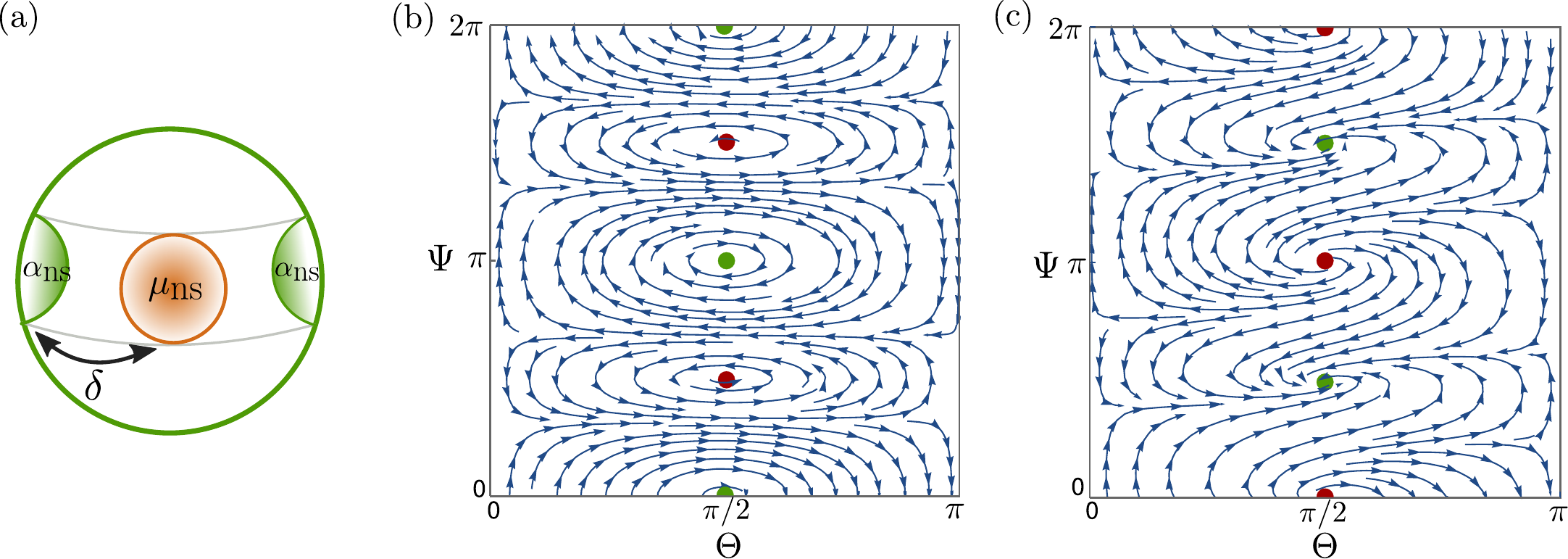}
	\caption{
        (a) An example of a colloid with non-axisymmetric coating described by Eq.~\eqref{eq:non-axi-coating}. (b,c)   Phase-space dynamics of the angles $(\Theta,\Psi)$, determined by Eqs.~\eqref{eq:theta} and \eqref{eq:psi}. In panel (b), the parameters have been set to $\lambda=0.1$ and $W = 0.25$, and, in panel (c), we have used $\lambda = -2$ and $W = -0.25$. The green (red) dots represent the stable (unstable) fixed points of the motion.}
	\label{fig:fig3}
\end{figure}

We now briefly discuss phoretically active colloids with non-axisymmetric surface coatings.
In order to design a swimmer with profiles that lead only to rotational motion, we consider the following surface mobility and activity
\begin{align}\label{eq:non-axi-coating}
    \alpha(\theta,\phi)=
    \alpha_{\rm ns}(1-\cos^2\phi)(1-\cos^2\theta),\quad
    \mu(\theta,\phi) = \mu_{\rm ns} [1-\cos^2(\phi+\delta)](1-\cos^2\theta),
\end{align}
where $\delta$ is the phase difference between the mobility and activity distributions [see Fig.~\ref{fig:fig3}(a)].
Due to the symmetry of the system, the expansion of activity and mobility in spherical harmonics only yields nonzero coefficients for even terms up to $\ell=2$. 
In particular, we find
\begin{align}
    A_{00}&= \frac{\sqrt{4\pi}\alpha_{\rm ns}}{3}, & A_{20} &= -\frac{\sqrt{4\pi}\alpha_{\rm ns}}{3\sqrt{5}}, &A_{22}&=A_{2-2} = -\frac{\sqrt{4\pi}\alpha_{\rm ns}}{\sqrt{30}},\\ 
    M_{00}&= \frac{\sqrt{4\pi}\mu_{\rm ns}}{3}, & M_{20} &=-\frac{\sqrt{4\pi}\mu_{\rm ns}}{3\sqrt{5}}, & M_{22}&=M_{2-2}^*= -\frac{\sqrt{4\pi}\mu_{\rm ns}}{\sqrt{30}} e^{i2\delta}.
\end{align}
Since only the indices $m =-2,0,2$ have non-zero coefficients, we can see from Eq.~\eqref{eq:Blm} and \eqref{eq:Dlm} that the terms with $\Delta_{1}^\pm(\ell,m)$ must vanish since they only involve odd $m$ indices, while those with $\Delta_{2}^\pm(\ell,m)$ can remain.

The stresslet tensor \eqref{eq:phS} can be calculated analytically, and the corresponding net angular velocity becomes $\boldsymbol{\Omega}=\boldsymbol{\Omega}^{\rm e}+\boldsymbol{\Omega}^{\rm o}$ where
\begin{align}
    \boldsymbol{\Omega}^{\rm e}
    &=
    -
    \frac{\alpha_{\rm ns} \mu_{\rm ns}}{15aD}
    \sin2\delta\,\mathbf{p},\\
    \boldsymbol{\Omega}^{\rm o}
    &=
    \frac{\alpha_{\rm ns} \mu_{\rm ns}}{105aD}
    \mathbf{e}\cdot
    \left[
    3(2+\cos2\delta)\left(\mathbf{pp}-\frac{1}{3}\mathbf{I}\right)
    +
    (8+\cos2\delta)(\mathbf{e}_1\mathbf{e}_1-\mathbf{e}_2\mathbf{e}_2)
    -\sin2\delta(\mathbf{e}_1\mathbf{e}_2+\mathbf{e}_2\mathbf{e}_1)
    \right]\cdot\mathbf{G}(\lambda)
    .
\end{align}

The swimmer spins with angular velocity $-\frac{\alpha_{\rm ns} \mu_{\rm ns}}{15aD}\sin2\delta$ and simultaneously exhibits rotation due to odd viscosity, which results from the stresslet moment.
Note that the stresslet tensor in general has three different eigenvalues, making the dynamics more complex than in the axisymmetric case discussed previously \cite{hosaka2024chirotactic}. 
Even when the surface modes have two-fold symmetry about $\mathbf{p}$, i.e, $\delta=\frac{\pi}{2}k$, the odd angular speed remains finite, whereas the even component along $\mathbf{p}$ vanishes.
For $\delta=\pi k$, the swimmer recovers the axisymmetric profile about the axis $\mathbf{e}_2$ and the velocity is solely due to the stresslet along this axis, $\boldsymbol{\Omega}
=-\frac{6\alpha_{\rm ns} \mu_{\rm ns}}{35aD}\mathbf{e}\cdot\left(\mathbf{e}_2\mathbf{e}_2-\frac{1}{3}\mathbf{I}\right)\cdot\mathbf{G}(\lambda)$. 

We next consider the special case where $\delta = \pi/2$. We define the odd viscosity direction to be $\mathbf{e}\equiv\mathbf{e}_z$ where $\mathbf{e}_z$ is the unit vector of $z$-direction in the lab frame. 
This allows us to obtain the equations of the Euler angle dynamics $(\Theta,\Phi,\Psi)$ of the swimmer where we follow the same convention as in Ref.~\cite{hosaka2024chirotactic}. The equations are then reduced to
\begin{align}
        \dot \Theta &=W\sin\Theta\big[4 g_2 (3+7\cos 2\Psi)\cos\Theta+28g_1\sin 2\Psi\big],\label{eq:theta}\\
        \dot \Phi &= W\big[12 g_3\sin^2\Theta - \lambda  -28\cos 2\Psi(g_1+g_3\cos^2\Theta)+28g_2\sin 2\Psi \cos\Theta\big],\\
        \dot \Psi &= W\big[4g_1(3+7\cos 2\Psi)\cos\Theta-28g_2 \sin 2\Psi \big],\label{eq:psi}
    \end{align}
    where $W \equiv \frac{\alpha_{\rm ns} \mu_{\rm ns}}{420aD}$ and $g_b = (\lambda/4)^b / [1+(\lambda/4)^2]$. As expected, the equations depend only on $\Theta$ and $\Psi$ due to the cylindrical symmetry imposed by the odd viscosity axis \cite{hosaka2024chirotactic, Khain2024Aug}. Depending on the sign of $W$ (i.e., the sign of $\alpha_{\rm ns} \mu_{\rm ns}$), there are only two stable fixed-points for the dynamical system $(\Theta^*,\Psi^*)$. As shown in Fig.~\ref{fig:fig3}(b), when $W >0$ the pairs of stable fixed points (green) are located at the points $(\pi/2 , 0)$ and $(\pi/2 , \pi)$ 
    and the unstable fixed-points (red) are located at $(\pi/2 , \pi/2)$ and $(\pi/2 , 3\pi/2)$. On the other hand, when $W<0$ [Fig.~\ref{fig:fig3}(c)], the nature of the fixed-points is reversed, which means that the stable fixed-points are located at $(\pi/2 , \pi/2)$ and $(\pi/2 , 3\pi/2)$ and the unstable ones are located at $(\pi/2 , 0)$ and $(\pi/2 , \pi)$. 
    In both scenarios, two additional saddle points are located at the poles of the $2$-sphere in which $\Theta^*=0,\pi$. 
    From our analysis, we can therefore conclude that in the long-time limit the particle will align in the plane perpendicular to the axis of odd viscosity regardless of the sign of $W$ and $\lambda$.

\section*{Concluding Remarks}

In this work, we have studied the motion of a self-phoretic colloidal particle moving in a 3D chiral active fluid with odd viscosity. For a spherical swimmer with arbitrary surface slip velocity, the generalized Lorentz reciprocal theorem has shown that the translational velocity~\eqref{eq:V} is completely independent of the odd viscosity.
In contrast, the rotational motion~\eqref{eq:Omega} is affected by odd viscosity and is associated with the stresslet moment of the surface mode. By calculating the stresslet of a spherical swimmer with the activity and mobility expanded in spherical harmonics [Eq.~\eqref{eq:phS}], we have determined the angular velocity of a swimmer with arbitrary mobility and activity coatings. As a showcase application of the general formulae, we have derived the velocity of a Janus particle and found that the chirality in the surrounding fluid induces rotational motion as well as translational motion, which leads to helical trajectories, followed by reorientation behavior (chirotaxis).
We have also discussed the effect of non-axisymmetric surface modes on the angular velocity and have shown that the general stresslet moment can lead to purely rotational motion.

Experimental realizations of fluids with odd viscosity have been performed predominantly in 2D systems, such as fluids with rotating granular gears~\cite{yang2021topologically} and with microscopic driven spinners~\cite{soni2019odd, lopez2022chirality, mecke2023simultaneous}; a notable exception is Ref.~\cite{chen2024self}. Our findings present a promising outlook for a new generation of experiments with self-phoretic active colloids in three dimensional chiral active fluids. This is because swimmers with any given slip velocity profile do not exhibit any form of activity related to the odd viscosity in 2D~\cite{hosaka2024chirotactic}, and the emergent rotational dynamics predicted here can exist only in 3D. 

It has been shown that dragging objects with asymmetric shape (such as a triangular object) can lead to rotational motion in a bulk fluid with odd viscosity, due to rotation-translation coupling in their full resistance tensor~\cite{Khain2024Aug}. Our findings have shown that even a spherical particle undergoes rotational motion once it becomes active, due to the coupling between the odd viscosity and the stresslet moment~\eqref{eq:stresslet}.

The expressions for the swimming velocity~\eqref{eq:V}, \eqref{eq:Omegae}, and \eqref{eq:Omega} can be utilized to describe the dynamics of other types of surface-driven microswimmers, such as electrophoretic~\cite{Ibrahim2017Oct,vanBuel2026Apr} or thermophoretic colloids~\cite{golestanian2012collective,PhysRevLett.112.068302}.
In all these cases above, the slip velocity is related to gradients of the corresponding scalar phoretic field as in Eq.~\eqref{eq:vs} and the resulting stresslet, which is required for determining the angular velocity, has been obtained in Eq.~\eqref{eq:phS}. 
For future investigations, our framework can be extended to non-spherical colloids~\cite{michelin2017geometric}, such as ellipsoids~\cite{Fair1989Feb} or slender bodies~\cite{Solomentsev1994Nov, ibrahim2018shape, golestanian2007designing}.
In general, active particles disturb the fluid around them as stresslets that govern their collective dynamics~\cite{lauga2016stresslets}.
Hence, it will also be of value to explore the collective behavior of active particles in chiral fluids within coarse grained (bottom-up) approaches~\cite{Tucci2024Jul} or numerical simulations~\cite{Brady2026} 
applied to Janus colloids.

\begin{acknowledgments}
We thank J.\ C.\ Everts for useful discussion.
We acknowledge support from the Max Planck Center Twente for Complex Fluid Dynamics, the Max Planck Society, the Max Planck School Matter to Life, and the MaxSynBio Consortium, which are jointly funded by the Federal Ministry of Education and Research (BMBF) of Germany.
Y.H.\ acknowledges support from the Japan Society for the Promotion of Science (JSPS) Overseas Research Fellowships (Grant No.\ 202460086).
A.V.\ acknowledges support from the Slovenian Research and Innovation Agency (Grants No.\ P1-0099 and No.\ J1-60009). 
M.C.\ acknowledges support from EPSRC through grant EP/Z534766/1.
\end{acknowledgments}

\bibliography{apssamp,Golestanian}

\end{document}